\documentclass[manuscript]{acmart}

\usepackage[utf8]{inputenc} 
\usepackage{graphicx} 
\graphicspath{ {./} }
\usepackage{multirow} 
\usepackage{soul, color} 
\usepackage{hyperref} 
\usepackage{url} 

\title{Data Quality, Mismatched Expectations, and Moving Requirements: The Challenges of User-Centred Dashboard Design}
\author{Mohammed Alhamadi}
\affiliation{%
  \institution{The University of Manchester}
  \city{Manchester}
  \country{United Kingdom}}
  \email{mohammed.alhamadi@manchester.ac.uk}
  \author{Omar Alghamdi}
\affiliation{%
  \institution{The University of Manchester}
  \city{Manchester}
  \country{United Kingdom}}
  \email{omar.alghamdi@manchester.ac.uk}
\author{Sarah Clinch}
\affiliation{%
  \institution{The University of Manchester}
  \city{Manchester}
  \country{United Kingdom}}
  \email{sarah.clinch@manchester.ac.uk}
\author{Markel Vigo}
\affiliation{%
  \institution{The University of Manchester}
  \city{Manchester}
  \country{United Kingdom}}
  \email{markel.vigo@manchester.ac.uk}
  
  \copyrightyear{2022}
\acmYear{2022}
\setcopyright{acmlicensed}\acmConference[NordiCHI '22]{Nordic Human-Computer Interaction Conference}{October 8--12, 2022}{Aarhus, Denmark}
\acmBooktitle{Nordic Human-Computer Interaction Conference (NordiCHI '22), October 8--12, 2022, Aarhus, Denmark}
\acmPrice{15.00}
\acmDOI{10.1145/3546155.3546708}
\acmISBN{978-1-4503-9699-8/22/10}

\begin{document}

\begin{abstract}
Interactive information dashboards can help both specialists and the general public understand complex datasets; but interacting with these dashboards often presents users with challenges such as understanding and verifying the presented information. To overcome these challenges, developers first need to acquire a thorough understanding of user perspectives, including strategies that users take when presented with problematic dashboards. We interviewed seventeen dashboard developers to establish (i) their understanding of user problems, (ii) the adaptations introduced as a result, and (iii) whether user-tailored dashboards can cater for users' individual differences. We find that users' literacy does not typically align with that required to use dashboards, while dashboard developers struggle with keeping up with changing requirements. We also find that developers are able to propose solutions to most users' problems but not all. Encouragingly, our findings also highlight that tailoring dashboards to individual user needs is not only desirable, but also feasible. These findings inform future dashboard design recommendations that can mitigate the identified challenges including recommendations for data presentation and visual literacy.
\end{abstract}

\maketitle

\section{Introduction}
\label{intro}
Interactive information dashboards enable the interaction with complex data sets using visualisations, tables or maps, often presenting multiple such widgets together on a single display. Dashboards are used in many domains such as healthcare \cite{buttigieg2017hospital, dowding2015dashboards, koopman2011diabetes}, education \cite{roberts2017give, schwendimann2016perceiving, jivet2018license} and urban development \cite{kitchin2015knowing, pathak2015city, lee2015cityeye}. Despite dashboards' increased adoption, users face interaction and information comprehension challenges caused mainly by information overload \cite{sarikaya2018we, buttigieg2017hospital} and visual literacy gaps~\cite{wakeling2015graph, srinivasan2018s, dowding2018impact}. These challenges are complex and multifaceted, but evidence suggests that they partially emerge from dashboard developers prioritising visual appeal over functional effectiveness \citep{few2006information, eckerson2010performance}.

Addressing the problems users face is key as dashboards continue to be crucial artefacts to aid decision making in important areas such as epidemiology and finance. Recently, in response to the COVID-19 pandemic, many interactive dashboards have been developed by worldwide health agencies, non-governmental organisations and individuals \citep{WHODashboard, COVIDView, NHSDashboard, SaudiDashboard}. These dashboards visualise epidemiological data to track disease outbreaks as they unfold, and monitor and report on future outbreaks \citep{dong2020interactive}. In response to existing problems, public health officials have been creating tools and tutorials to encourage users to become dashboard co-creators, think about data modelling and manipulation and to critically assess the making of COVID-19 visualisations. A project such as \textit{We Rate Covid Dashboards} \citep{weRateCovidDashboards2021} reflects the magnitude of the problem: it rates COVID-19 dashboards taking into account visual presentation, navigation and data detail. No more than 25\% of 369 reviewed dashboards received a high rating, with only $\sim$1\% receiving the top rating.

Although a body of research has been conducted on user-centred dashboards \citep{dowding2018impact}, users still encounter problems. Since the lack of user involvement during development can result in dashboards that are difficult to use~\citep{roberts2017give, lee2015cityeye}, dashboard developers should consider the different levels of visual literacy of the intended users, the constraints of the layouts and the information presented to promote engagement and trust \citep{srinivasan2018s, herder2020privacy, wakeling2015graph, kia2020patterns}. In a similar line, previous research examined software developers' awareness of accessibility and how they can achieve success in mobile application and game development~\citep{antonelli2018survey, srisopha2021should, kultima2015developers}. However, aside from the scarce work demonstrating the challenges of user-centred dashboard development \citep{dimicco2016user}, the role of developers around these challenges and their solutions is yet not well understood. Hence, it is imperative to capture dashboard developers' perspectives in order to determine if dashboard development is inherently difficult and whether developers are aware of the problems users encounter. Thus, our work aligns developers' understanding of dashboard problems that are reported in the literature with appropriate solutions for addressing them.

Rather than pushing users into developing workarounds to overcome the problems they encounter, recent research has suggested that dashboards should cater for the individual users' characteristics  \citep{peischl2013can, sarikaya2018we, weggelaar2018developing, dabbebi2017towards, vazquez2020representing, eckerson2010performance}.
Three potential mechanisms for doing this are customisation, personalisation and automatic adaptation \citep{vazquez2019tailored}. We define any technical intervention introduced by developers to mitigate a problem an `adaptation', which may include the following tailoring mechanisms: `customisation' is performed by the user to change any aspect of the dashboard to their needs while `personalisation' is done by the system at dashboard creation or loading time. `Automatic adaptations' (not to be confused with the previous `adaptation') are real-time updates to the dashboard based on the user models. We refer to the use of one or more of these three mechanisms on dashboards as \emph{tailoring}.

In this paper, we present the findings from interviews with seventeen dashboard development experts. Our interviews explore their awareness of user problems that have been reported in the literature, the development practices contributing to these problems and the possible solutions to them. Our specific research questions are:
\begin{itemize}
    \item \textbf{RQ1.} What are the dashboard users’ \emph{challenges} from the developers' point of view? Are developers aware of the existing interaction problems?
    \item \textbf{RQ2.} What are the possible \emph{adaptation techniques} to the challenges? Can tailoring, especially \emph{automatic adaptations}, solve these challenges?
\end{itemize}


\looseness=-1
Our contributions are the following: 
\begin{itemize}
    \item We learned that developers are well aware of most of the literature-reported problems, as there is a considerable overlap between what developers report and the literature on dashboard problems -- the overlap is not complete though.
    \item We identified new user problems' with dashboards, as reported by developers, which were not in the literature such as users' tendency to verify the displayed data.
    \item We isolated the causes of known problems: making sense of the data is one of the most frequent problems users have, which is largely caused by a gap in the visual literacy users possess and the one developers expect. We also learned about key development practices that exacerbate existing problems including the pressure to deviate from design guidelines, which leads to ineffective data presentation issues.
    \item Finally, our findings suggest the feasibility of automatic adaptations to dashboards in order to address users' interaction problems as demonstrated by the adaptations suggested and employed by the developers.
\end{itemize}




\section{Related Work}
\label{2}

In this section we present first a literature review on the existing barriers encountered by dashboard users. Then we discuss the mechanisms for adaptive dashboards reported in the literature, which do not necessarily correspond to the mentioned barriers.

While information dashboards are ubiquitous nowadays, understanding and using dashboards has always been problematic. Existing difficulties are sometimes attributed to a gap between the visual literacy developers expect and the actual one~\citep{peer2019community, vornhagen2018sensemaking, sarikaya2018we, weggelaar2018developing}. This gap is widened in domains that are inherently complex such as city governance \citep{vornhagen2018sensemaking} or with users with low visual and analytic literacy \citep{weggelaar2018developing, barnett2019digital}. Difficulties in understanding data on dashboards can also be caused by poor information presentation \citep{sarikaya2018we}. Data representation using graphs can be described as \textit{dry} leaving some users uninterested \citep{hagood2016integrating}. When it comes to self-tracking data, presenting correlations without overwhelming users with data is a known challenge \citep{jones2015exploring}. For users who need to know which information is relevant for decision making, some authors demand explicit instructions to operate dashboards~\citep{colley2016insights}. Another example of misalignment highlights the lack of correspondence between the data and users needs in learning analytics dashboards \citep{echeverria2018driving}.

Data sharing, privacy and security have also been common issues in healthcare \citep{tendedez2018scoping, cohen2017impacts} and education-related dashboards \citep{haupt2015using, yoo2019designing}. Both academic advisors and students have concerns about sharing student data voluntarily even if it is intended to support student mental health or academic achievement \citep{yoo2019designing, sun2019s}. Handling fragmented or incomplete data sources is another frequently reported problem where siloed data (data fragmented across different information systems) is detrimental for decision making \citep{haupt2015using}. In healthcare settings, clinicians are reluctant to act on dashboards if they do not have access to the source of the data and the meta-data, which indicates a lack of trust on dashboards~\citep{tendedez2018scoping}.

Another challenge comes from the inefficiency of support for user-requested features such as adjustments to the granularity of data aggregation and the functionalities to enable comparison, customisation and annotation tasks \citep{sarikaya2018we, elias2012annotating}. Also, user training was described as a \textit{huge} challenge especially for clinicians who are usually time-constrained \citep{tendedez2018scoping}. On the other hand, some works also investigate challenges in implementing dashboards \citep{cohen2017impacts, haupt2015using, sarikaya2018we} including data quality and tools' cost. Data quality problems in dashboards have negative consequences for data completeness, consistency and accuracy \citep{cohen2017impacts}. Other authors report that business intelligence tools used to create dashboards are complex and need constant IT support \citep{haupt2015using}.

Dashboard design has also been an area of interest with works leading to design goals \citep{hagood2016integrating, charleer2018real}, design proposals \cite{echeverria2018driving, jones2015exploring} and design implications \citep{tendedez2018scoping}. Design goals include adaptability, transparency, intelligence and glanceability \citep{charleer2018real}, while design suggestions include using a data storytelling approach as a learning design \citep{echeverria2018driving}. \citet{jones2015exploring} suggests comparing self-tracking data based on the user's needs or using feedback from other users. Involving all stakeholders in the dashboard design and development process has also been explored in education and operational monitoring with encouraging results and positive impact \citep{gilliot2018participatory, martins2017development}.

\looseness=-1
Recent research recognised that the full potential of dashboards could only be realised if individual users' characteristics are factored into design \citep{peischl2013can, sarikaya2018we, weggelaar2018developing, dabbebi2017towards, vazquez2020representing}, which is an explicit call to user-centred design of dashboards. Some authors suggest this could be achieved by providing support for customisation to adapt dashboards to users' needs~\cite{sarikaya2018we}, where automatic adaptation to different users and displays is still an open research problem. This idea was supported in the context of increasing dashboard understanding through the use of adaptive functionalities \cite{weggelaar2018developing}. Some model-driven approaches to automatic adaptation include adaptive learning analytics dashboards based on users' activities, preferences and objectives \cite{dabbebi2017towards} and a visualisation recommender system based on use context, user characteristics, domain and tasks \cite{vazquez2020representing}. 

\looseness=-1
Our approach goes beyond the identification of dashboard problems: we firstly explore developer awareness of the aforementioned user problems and elicit developers' views on usable adaptation techniques to lessen existing problems. Subsequently, we look into the feasibility of tailoring dashboards as a way to inform future interventions. Thus, our paper contributes to (1) understanding developers' awareness of users' problems with dashboards and contrasting them to known ones as reported in the literature, (2) isolating new user interaction problems, (3) identifying the reasons behind the challenges from a developer perspective and (4) assessing the feasibility of using automatic adaptations to address users' interaction problems. While we find developers to be aware of these issues, users still encounter problems. This suggests that developer awareness alone is not sufficient and further interventions are required in the development process of dashboards.

\section{Method}
Dashboard development experts were recruited from specialised online forums including Tableau Community, Microsoft Power BI Community and Sisense Community. Thus we used purposeful and snowball sampling techniques. Participants were eligible if they had been working on analysing, designing or developing dashboards or visualisations for dashboards for at least two years, and if the dashboards were created for someone else to use. Semi-structured interviews were conducted remotely using a video conferencing tool and participants were incentivised with a £15 Amazon voucher. Ethical approval was obtained from The University of Manchester Research Ethics Committee.

\subsection{Interview Topics and Materials}
Our interviews explored the following topics: (i) experience and training undertaken to understand and work on data and dashboards; (ii) problems reported by users; (iii) techniques used and suggested to address reported problems; and (iv) tailoring dashboards to users (customisation, personalisation and automatic adaptation). To illustrate what we meant exactly by adaptations, we shared with developers six pairs of challenges and adaptations from the literature in textual format via a website (available in the supplementary material). Examples include customisation support by incorporating functionalities such as drag-and-drop, and recommending the most appropriate visualisations based on the users' goal, inferred by monitoring user behaviour and building user models. In the same way, we also showed developers adaptations on web pages \citep{brusilovski2007adaptive} to get their view on whether similar adaptations could be applied to dashboards (available in the supplementary material). Examples were: (1) \emph{page adaptation} to select the most suitable version of a web page according to an interaction context model, and (2) \emph{adaptive navigation} to generate, disable or alter the appearance of hyperlinks based on the goals, preferences, and knowledge of users.



\subsection{Participants}
We recruited seventeen participants (13 males and 4 females) from the UK (n=$5$), USA (n=$5$), India (n=$3$), Australia (n=$2$), Portugal (n=$1$) and Malaysia (n=$1$). The average age of the participants was 39 years old (SD =$10$). Participants had on average $9.2$ years experience developing dashboards (SD=$5$), and the majority worked in the private sector (n=$13$), while others were in the academic (n=$3$) and public sector (n=$1$). Participants had a high education level (one held an associate degree, eight held bachelors degrees, and eight higher degrees) and came from many backgrounds such as finance, business management, psychology, history, mathematics, computer science, information systems and engineering. Six participants were independent contractors while eleven worked on dashboards for their organisations.

Participants had been involved in the development of operational (13), strategic (10) and analytical (7) dashboards~\citep{few2006information} along with other types that participants categorised as quality, clinical, reporting, tactical, prediction, dynamic, efficiency, process-oriented, informative and executive dashboards. These dashboards were used in diverse settings such as financial (9), healthcare (8), telecommunication (4), sales (3), business (2), IT (2), among others. Participants used a plethora of tools to create dashboards with Tableau (15) being the most popular followed by Power BI (5), Alteryx (2) Oracle OBIEE (2), Qlik Sense (2), Cognos, Microstrategy, Shinyapps, Spotfire, Qlikview, Salesforce and SAP BusinessObjects. Tableau and Power BI constituted the majority of the tools (45\%) confirming existing evidence on dashboard development platform uptake~\citep{Richardson2021}. Most developers had experience with a combination of these tools.

\subsection{Analysis}
It took an average of 68 minutes to conduct the interviews (SD = 13 minutes) which were audio recorded and then transcribed. Dedoose 8.3 was used to perform inductive thematic analysis following \citeauthor{braun2006using}'s methodology \citep{braun2006using}: data familiarisation, code generation, themes search and revision, themes definition and naming, and then report production. We first analysed dashboard problems in an emergent fashion to answer RQ1. Then, to answer RQ2, we followed a-priori coding to assign adaptations to tailoring mechanisms from the literature (customisation, personalisation and automatic adaptation). A co-author who was not involved in the project was given the transcripts and codebook to test the reliability of the coding. The inter-rater reliability results showed a moderate agreement, Cohen’s $\kappa = 0.66$.

\section{Results}
\label{4}
We report results of our analysis in two subsections. In Section~\ref{sect:emerging}, we thematically analysed the responses and assigned a total of 61 codes which yielded five main themes. The themes that emerged from our thematic analysis include a set of challenges that are written \textbf{in bold}. Data in Section~\ref{sect:apriori} was analysed following a-priori coding, where the classified instances within the categories are extracted from the literature (customisation, personalisation and automatic adaptation). Tables 1--4 in the Appendix section provide a mapping between the challenges, adaptations proposed by the developers and implementation strategies suggested by the authors.

\subsection{Dashboard Challenges}
\label{sect:emerging}
\subsubsection {Involving Users in the Development}
Most developers (n=12) involve users throughout the dashboard development process while the rest involve them intermittently. In the latter case, users participate at the outset for requirements elicitation purposes, at the end to summatively evaluate the dashboards or at both stages. It is agreed that collecting requirements from end users is a more effective practice than involving other stakeholders such as IT staff. While this may seem obvious, end users are more knowledgeable about their abilities and limitations and will have a clearer understanding of how the dashboard can, or cannot, serve their needs. Developers seek consensus on the requirements although the number of users and their availability prevents this from happening often times. As the number of users involved grows, the more diverse (often contradicting) opinions have to be considered. Most users require training but the individual differences makes training many users unfeasible. These differences include users' visual literacy, their willingness to adopt a new technology and the perceived benefit gained from such adoption \citep{yera2019modelling, yigitbasioglu2012review, weggelaar2018developing}. Finally, arranging meetings with users in senior positions is a challenge in itself.

\textit{P16: ``the ones that tend to be less agile are the ones where there are more departments involved because the more people you get involved, particularly senior staff, the harder it is to pin them down into meetings and to have consensus because we try and only have one dashboard that they all agree with.''}

Prior to starting the development, developers agree on the need to answer questions such as `why is this dashboard being developed?', `what will be conveyed?' and `who is the user that will be told the story?' This information helps developers to design the dashboard so it conveys the intended story. Developers often misunderstand user needs as defined in the requirements, which leads to a misalignment between what the users need and what the developer implements (\textbf{implementation misalignment}). When collecting requirements, developers notice that users struggle to articulate their needs or do not have a clear idea of what their needs are. To overcome this problem, users are given a list of KPIs to trigger their thinking on their needs.


Most developers (n=15) use dedicated software tools to build their dashboards. The majority of the tools are proprietary and very few developers use open source tools, which are typically more portable and extensible. Prior to development, some developers conduct workshops with users to understand what tools users have or can afford. If requirements cannot be supported with these tools, feature requests are submitted or an investment in another tool is made. As a result, some design decisions are dependent on the tools used. For example, the collected usage data is limited to what the platform shares with the developers, while customisation support is entirely left to the tool.

\textit{P15: ``Not on Shinyapps at the moment, but certainly on Tableau, it comes with the service. We actively monitor who uses the Tableau server and who returns every month to show that it's got value.''}

The majority of interviewed developers (n=14) collect dashboard usage data, which is mostly provided by the tools although sometimes collection is done offline by speaking to users directly. This data is used to find interaction patterns and monitor performance. The collection frequency varies from monthly, every six months, to passive collection to be analysed only if there is a reported issue. Also, some developers collect data out of personal interest and not for any specific purpose.
\looseness=-1

\textit{P16: ``I get nosy sometimes and I want to see who is viewing the stuff that I have built because I want to know that it is still relevant to the business, sometimes it is and sometimes it is not.''}

Most developers (n=10) say that users report the problems they face with dashboards. An absence of reports is understood as a lack of problems, which is mainly due to the adjustments made to dashboards responsively in existing feedback loops, or because users are taken along in the development process and all their needs are catered for.

\textit{P1: ``I think if they are designed properly and they are co-produced with users and you take into account their needs and you listen to what they are saying then actually you make them usable, so then they do not have any issues.''}

\subsubsection{All about Data: Performance, Access, Provenance, Currency and Metadata}
\label{4.2}
\looseness=-1
Data is not always available or requires further processing so it can be usable (\textbf{data quality issues}). Data unavailability is exacerbated when the users are responsible for storing their own data and when the data is fragmented and siloed across different information systems. Moreover, developers report that \textit{when}, \textit{how} and \textit{where} data was entered is important for decision making, which can also be problematic when users are in charge of collecting and sharing their own data (mostly not captured). In addition to provenance, the currency of the data is another user concern. Users usually complain if the data shown on a dashboard does not refresh at the users' required rate. Since developers are aware that the amount of data they pull in determines how quick visualisations render, requests for more granular data puts at risk the performance of the dashboard.
    
\textit{P4: ``Performance is very key to any kind of reporting. No one wants to be stuck on a graph or an object in the report rendering for more than five or ten seconds. If you are going to a granular level of data that is a different thing, it might take a little time because lots of rows need to be pulled in.''}
    
Developers employ data caching techniques to improve \textbf{dashboard performance}. Another technique is to load data at the background that may be closely related to the current, to be displayed in case it is requested on demand. When displaying large data sets that the dashboard cannot handle, a data extract (a compressed version) or a periodic refresh is performed instead of showing live data. For data from heterogeneous sources with different formats, developers use dedicated tools to handle them.


Developers receive a high number of reports from users when dashboards do not show the expected information (\textbf{data verification}), which is often due to the display of data the users were unaware of. When users are not sure of the accuracy of the information presented, they usually question the origin of the data. If the dashboard does not capture the depth of the data that satisfies users' needs, users check the raw data sets frequently. Developers try to prevent this workaround as they believe it undermines dashboards' value of providing data overviews.

\textit{P14: ``a major challenge I've faced across multiple clients is the data mismatch. An executive, for example, sees everything that we have on Tableau. If they look at different views that we have created specifically for marketing they will say this data does not look right to me because I see my numbers on a daily basis, my number is 500 whereas on the marketing dashboard I see the number is 400. What they don't realise is that the approaches that we had for two different dashboards in the back end can be different or there can be filters on the default view to cater for the need of that specific function.''}
    
Building users' trust in analytics is problematic when the presented information clashes with users' biases (\textbf{lack of trust}). Developers agree that showing more information about the source of data and metadata can help users trust the data more. Also, allowing users to zoom into the data set can build trust in the analytics because, for example, they have a better view of how data was aggregated. It was also suggested that building trust in the developer and the tools used increases the overall trust on the dashboard.
	
\textit{P12: ``trust in the analytics only comes from allowing quite a deep dive, especially with the clash between the gut instinct and what somebody sees within the data set [\ldots] Quite often, it’s a much longer term journey than just one piece of analytics or one dashboard to build that trust.''}

\looseness=-1
Developers' efforts to create barrier-free dashboards is challenged by issues such as \textbf{interoperability} and authentication processes needed to \textbf{protect data access} and personalise user experience. To make sure that users have access privileges to the data requested, developers use workflows connecting data sources to user roles. Interoperability problems, at the data or platform level, within organisation or across organisations makes the integration of dashboards into users' daily life difficult.

\subsubsection{The Tensions of Addressing User Needs}
\label{4.3}
Dashboard users ask for new data frequently so developers find it difficult to fit all the required information on a single screen that users can grasp instantly. Putting too much information on a single display leads to \textbf{information overload} and prevents effective decision making. As a result, developers typically push back on those requests. To balance users' frequent requests for more data, developers suggest combining data into smaller subsets to then enabling drilling into more granular data on demand \citep{shneiderman2003eyes}. To reduce the number of charts in dashboards, users are encouraged to keep those charts that are still relevant and frequently remove those that are not needed anymore.
	
\textit{P5: ``users ask us [to] add this also, add that also in that particular chart, so many times we need to educate them, yes we can have this but on a different dashboard, let us not put everything in one dashboard otherwise it will be more and more complex and difficult to understand.''}

Sometimes, by the time developers create the dashboard, users realise that the original problem has changed and the created dashboard no longer answers their question. Developers understand the importance of capturing evolving realities in the domains of interest, but find it difficult to keep it up with constant change. This can lead to dashboards that do not fully address all users' needs.

\textit{P12: ``Even the traditional agile two-week sprints don't work with modern dashboard building mostly due to the fact that you are highlighting and showing information visually, that very quickly people can go and pick apart, learn something new and ask another question that is not in the original scope.''}

Most participants (n=12) say they follow data visualisation guidelines when making the dashboards but some do not (n=5). The sources vary between literature, dashboarding tools’ best practices (e.g., Tableau \citep{Tableau2022}) and their own guidelines accumulating from experience. Some also find \emph{inspiration} on the Web (e.g., Google Image Search). Following guidelines is not always possible since users preferences can contradict best practices so there must be some ``flexibility.''

\textit{P14: ``Initially we tried following guidelines but users were not aware of whether this is the right approach or not and they were used to Excel so they wanted those Excel tables in Tableau, that's it!''}

\looseness=-1
Developers are sometimes asked to stick to specific chart types or colours because of the arbitrary preferences of managers even if they do not conform to the best practices (\textbf{ineffective data presentation}). Also, while users' visual literacy is taken into account, it can be challenging to accommodate all users as not everything can be represented with simple charts \citep{sarikaya2018we}. Moreover, to present the data effectively, developers need to understand the KPIs of a variety of industries and domains.  

\textit{P12: ``I absolutely take into account what they are used to using and what they are capable of using based on their data literacy and then from there it is using the right charting for the job as well. So that depends on how hard I have to fight them on that or how much I need to train them out to be able to use the tool effectively.''}

Developers agree that choosing visualisations should be informed by the questions that the dashboard is trying to answer. When a user asks for a specific chart that goes against the best practice, developers try to educate the users on the best way data should be presented. If the user is not convinced, some developers implement two charts on the dashboard and enable swapping between them. To support users in locating specific data, developers use filters on a dashboard (at their end), to then share that view with the users. Also, developers create responsive dashboards so they can be used on multiple devices.
When users need to use multiple dashboards, they spend more time to locate the needed information if the design is not consistent across the dashboards and when they need to navigate between them. They can also face visual exhaustion if the dashboards do not use colours properly, use excessive ink to present the data or is in an uncomfortable layout (e.g., not supporting access from different devices)~\citep{few2006information}.

\looseness=-1
\textit{P13: ``colours not being used properly, too much ink being used, not structuring the dashboard, and KPIs should be defined usually around the top of any dashboard you create… and standardisation is also another thing so it's more on using the right visuals…''}

\subsubsection{Adoption, Onboarding and Facilitating the Learning Journey}
\label{4.4}
Although users can be excited to use dashboards, conveying information with visualisations may be difficult because some users are used to traditional tabular reports. Even if the dashboard is made very simple with minimum interactive features, users may not understand what the dashboard shows or why certain charts are displayed. Developers also report that some users show willingness to learn but there are not enough learning resources or they do not know where to find them.
		

If the dashboard does not match users' visual literacy it hinders the sensemaking process and understanding the dashboard's purpose becomes difficult \citep{sarikaya2018we}. Developers believe that users need to learn how to interact with dashboards, for example, how to hover over objects to find tool tips and how to use menus. To make matters more difficult, developers' creativity often affects further the comprehension \citep{galesic2011graph}.
    
\textit{P1: ``I was quite concerned that in the visualisation world, people can produce quite sophisticated visualisations of data but in the clinical world the people that I work with wouldn't necessarily understand that at all. People's understanding of data especially in front line clinical practice is not that sophisticated, so you need to make sure the data you are displaying they understand what it means.''}

The more interactivity and advanced features a dashboard has, the more complex it becomes. So training the users to use the dashboards becomes challenging especially if they have different levels of visual literacy.
	
\textit{P9: ``Training is a big challenge because you have different people with different levels of knowledge of data, so someone can see a bar chart and understand exactly what it means, but a different person for the same use case sees a bar chart and takes hard time to understand what is the y-axis and what is the x-axis.''}

Developers believe that many users rely on assumptions rather than actual data and it is hard to change their mindset into data-driven decision making. Moreover, if the dashboard is a top-driven initiative (enforced by management instead of demanded by users) then users may not be inclined to use it and they will prefer more traditional ways of exploring data such as spreadsheets. On the other hand, if it is driven by the users themselves they become more interested and engaged. Additionally, dashboards that compare users performance to other users cause them to be \textbf{demotivated} and less likely to continue using them unless advice to improve their performance is included.

\textit{P11: ``auditing feedback dashboards instantly will say\ldots okay 70\% of your peers are better than you at this and that's great, but unless I actually know what to do to improve that's just going to make me feel not very great.''}


To mitigate visual literacy problems, developers typically train the users through a ``hand holding'' period to slowly guide them in early stages, where they go through the interpretation of visualisations several times. If there are different dashboards, users are put into groups and get introduced to their relevant dashboards. When \textbf{training users}, developers emphasise the capabilities of the dashboards and also their limitations. To enhance training within the dashboard, developers show users suggested actions, then ask for feedback on the suggestions.

\textit{P11: ``When we give people suggested actions we say we think you should do this because of X, Y and Z, then they have a mechanism to either agree with it and then it gets put on their personalised action plan or they can disagree and give a reason why, so we get feedback from the users and then it can iterate and incorporate into future designs.''}


Developers insist on including instructions which are typically shown in pop-up messages, videos, tutorials, tips and annotations. They mainly focus on the functionalities (e.g., filters) and dashboard navigation. Often, tutorials are then hidden but still made available whenever needed or shown to first time users or ones who have not logged in for a while. Developers aim to make dashboards easy enough for beginners to learn while robust enough to serve seasoned users' analytic purpose (\textbf{functionality use issues}).



To aid users in \textbf{understanding data} on dashboards, developers add extra helping information such as axes labels, legends or contextual information including data definitions, data reading strategies, description of visualisations and the purpose of the functionalities and the dashboard as a whole. Also, data points are suggested to provide enough context on maps and terms so developers do not depend on the users' knowledge.
	
\textit{P15: ``having a title that clearly states what the dashboard is showing so that users don't have to think about it is important, a description (from our point of view), an idea of the provenance of the data, what should we be looking at and what the point of the dashboard is...''}

It was also advised to use the least number of visualisation types so the users can understand all the charts (one or two types only). Developers use the same colours for the same kind of data across dashboards to enable users to distinguish data quickly (e.g., green for sales data, blue for financials, etc.). The smoothness of the flow and the arrangement of visualisations on the dashboard is important for the user to grasp what the developer is trying to convey. Choosing where to place visualisations and filters has a direct effect on the usability of the dashboard (\textbf{inappropriate placement}). Using white space to emphasise which charts are grouped together was an adaptation employed by several developers, which is in line with Gestalt principles of visual perception \citep{yigitbasioglu2012review}. Also, keeping common functionalities (e.g., filters) in one particular area in all of the dashboards makes it easier for users to know where they are and what they are for more quickly \citep{shneiderman1997designing}.
	
\textit{P11: ``when we had information on the right hand side of the page it was often missed by people so it was almost like they were reading the screen left to right so they'd see the main thing on the left and just work with that bit and wouldn't interact very often with the actions, whereas when we moved that box to the left-hand side \ldots people interacted with it a lot more.''}

\subsection{User-Tailored Dashboards to Improve Experience}
\label{sect:apriori}
This subsection describes how developers' implement customisation, personalisation and automatic adaptation on their dashboards. Users customise their dashboards while developers perform personalisation and automatic adaptations. Developers indicate that tailoring targets data, visualisation types, functionalities, layout and interaction.
\noindent \paragraph{\textbf{Customisation}}
More often than not, developers enable users to customise their dashboards (n=9). Those developers not providing customisation features express that allowing users to change some dashboard aspects can have negative consequences such as inaccurately representing data or the possibility of affecting other users who did not demand such features. As an intermediate solution, customisation features are often made available for advanced users only. Examples of implemented customisations include:
\begin{itemize}
    \item Enabling drilling down into more granular levels of data and sorting contents so that users can look at different outcomes and measures.
    \item Changing visualisation types (line graph or bar graph) and selecting the information they want to see.
    \item Enabling drag and drop to add their own tables or visualisations.
    \item Building dashboards through an interactive dialogue.
    \item Enabling users to select from a range of KPIs to show based on categories with the ability to go back to the KPI list and select others.
\end{itemize}

\textit{P8: ``I'm favourable to customisation but it depends on users' knowledge in terms of the data because if I give the user a bar chart and the option to change it to a gauge chart, some users will lose focus on the information because they would be putting information in a visualisation that is not suitable. So it's a plus if they are able to make those changes because in the end we are building dashboards for people to use in their daily work so they need to customise it to their needs.''}

\paragraph{\textbf{Personalisation}}
Personalisation is the form of tailoring used the most (n=13). Developers who do not implement personalisation features reported that it is not needed since all users' needs are already taken care of during development. Other developers do not implement personalisation features because they want to have full control of visualisations and interaction in order to evaluate the dashboard usability. Mostly, dashboards are personalised based on user roles and profiles. Examples on role-based personalisation include: changing filters to fit the role's needs, changing displayed information (summary for executive-level users then going deeper as more analysis is needed), and highlighting and restricting information relevant to that role. Another approach taken is by loading all the data in the background and enabling the user to switch to their relevant view (information and functionality). The last approach used is to store the state of the dashboard view after it is customised by the user for later use.

\textit{P14: ``an executive has 20 people in their team, including five managers and people reporting to them, the person at the bottom of the hierarchy only needs to view their own data, they cannot view data for their peers. Each manager can collectively view the data for who reports up to them, and for all of the managers combined the executive can view all data, so whenever a specific user logs in you define their hierarchy and the dashboard automatically restricts the data to that belonging to that user.''}

\paragraph{\textbf{Automatic Adaptation}}
Automatic adaptation is the most advanced form of tailoring since it alters the dashboard in real-time based on a user model -- only four developers reported using automatic adaptations. Those not employing adaptation techniques indicate that these features are too sophisticated, there are no tools to implement them, and ultimately users do not demand it. Examples of these automatic adaptations implemented include:
\looseness=-1
\begin{itemize}
    \item If it is detected that the user is facing problems, a tutorial is shown to them.
    \looseness=-1
    \item If there are erroneous elements in the data set, they will be taken out or dealt with before data is shown to the user.
    \looseness=-1
    \item Presenting information based on what motivates the user. For example, clinicians motivated by competition are shown percentage of patients at risk compared to the average across their group of users. On the other hand, clinicians motivated by the difference they perceive to make by their work are shown a chart of their performance over time.
    \item When a user requests a number of variables, the developer will add more variables than needed to accommodate probable future data needs.
    \item Recommending charts that are appropriate for the KPI that the user is trying to represent.
\end{itemize}


Developers generally agree that adapting dashboards to users is feasible and sometimes strongly advisable. They expressed several benefits to automatically adapting dashboards such as increasing interest by removing superfluous interactions, mitigating visual literacy problems by addressing users differences, and building trust by showing more relevant information.

\textit{P3: ``I love the idea, I believe it's delivering the information that's relevant to the user, [\ldots] I think it saves a step for the user of having to click on that, so I think that should be a focus and a goal of creating dashboards.''}

All of the developers agreed that adaptations such as the web-based ones shown to them can be used on dashboards. Some developers acknowledged they had been using page adaptation techniques on their dashboards. They shared the techniques to implement these adaptations such as adjusting how visualisations are displayed and how they can be operated depending on the screen size, and changing visualisations order based on the user's role and thus priority. Yet, they were also cautious and expressed some concerns: the complexity of inferring users characteristics from their behaviour and the importance of keeping data coherent so that different views can lead users to the same understanding.

\textit{P11: ``Definitely, this is all possible, [\ldots] the issue is when you don't know much about the user then you've got to try and infer it from their behaviour \ldots you can ask them to tell you what their preferences are or what their role is, or you have to infer it from their usage \ldots [which is] quite tricky.''}



\section{Discussion}
\subsection{Discussion of Findings}
We revisit the research question we formulate at the outset and discuss the implications on dashboard design:\\

\noindent\textbf{RQ1. What are the dashboard users' \emph{challenges} from the developers' point of view? Are developers aware of existing interaction problems?}\\ 
Our findings indicate that by involving users directly, developers become aware of the many problems users encounter with dashboards. However, when users do not report issues to developers, the perceived idea is that there are no problems. Since the absence of evidence is not evidence of absence, further research is needed to foster the user-developer communication effectively. 
Also, our results suggest that there are barriers that prevent effective communication between developers and users. Ineffective communication occurs before, during and after development. Before development, developers find difficulties to elicit user needs and establish their abilities. During development, developers struggle to accommodate frequent changes and requests for more data. After deployment, the challenge is in reporting user issues and mitigating them. This is exacerbated when developers do not collect and analyse usage data or when they collect it passively.

The most frequent problem is that users' visual literacy does not align with that of the artefacts and visualisations in dashboards, which are typically complex and sophisticated. Other problems include ineffective data presentation, data quality issues, functionality use issues, lack of trust and finding the motivations to continue to use the dashboards. While many of the problems are known from the literature~\citep{sarikaya2018we}, some other such as interoperability, data verification and implementation misalignment have not been reported in the literature. Moreover, developers do not seem to recognise some problems as reported in the literature such as user adaptability \citep{tokola2016designing, mazumdar2014exploring} and the lack of support for user-requested features (e.g., supporting comparison)~\citep{yigitbasioglu2012review}. We also provide a new perspective on known problems: when it comes to information overload, the fact that users demand more data makes it difficult for developers to control information overload. Our findings inform not only the problems and needs of dashboard users, but also those of developers when trying to address user needs. When involving users in the development, the most salient problem developers face are the frequent changes in the requirements and user demands.


Some users do not meet with developers directly but send intermediate parties (e.g., a personal assistant to a CEO) instead to negotiate their needs. This was not an anecdotal but a generalised practice that led to unmet expectations from the users. Even if users are involved, developers do not always include customisation functionalities if end users have low visual literacy. They claim that customisation will lead to non appropriate visualisations. When users are unable to articulate their needs there is little room for discussing the content users would like to have. Thus developers are selective about the user demands since some are perceived to be detrimental to the user experience. This tension in addressing some user demands results in creating suboptimal dashboards.

While the above-mentioned problems exist on their own, they do not always occur in isolation as one problem can trigger other problems.  The framework in Figure~1 illustrates the relationships between challenges in that existence of some challenges exacerbates the presence of others (\textit{may lead to} relationship) and fixing some challenges may generate others (the \textit{it may cause} relationship). For example, users with low visual literacy (who face data understanding issues) encourage developers to simplify dashboards (sometimes oversimplifying), making advanced users explore raw data to answer their questions, and inhibiting dashboards' potential of viewing data at a glance \citep{few2006information}. Alternatively, developers resort to training users on using the dashboards either in dedicated training sessions or within the dashboard as described in Section~\ref{4.4}. Often, this training is not only needed for users with visual literacy problems, but also for those who, despite having the skills, find it difficult to parse the intentions of developers. While this training is portrayed inevitable by the results in Section~\ref{4.4}, standardising domain-specific design guidelines and rigorously following them could reduce the need for it. Some other times, addressing one problem can create new ones: the findings in Section~\ref{4.2} indicate that trust on the dashboard can be built if up to date, low granularity data is provided, which is detrimental to dashboard performance and purpose. Developers are well aware of these effects and users develop workarounds when this information is not available.

\begin{figure*}
\label{fig:challenge_relationships}
\centering
\includegraphics[scale=0.35]{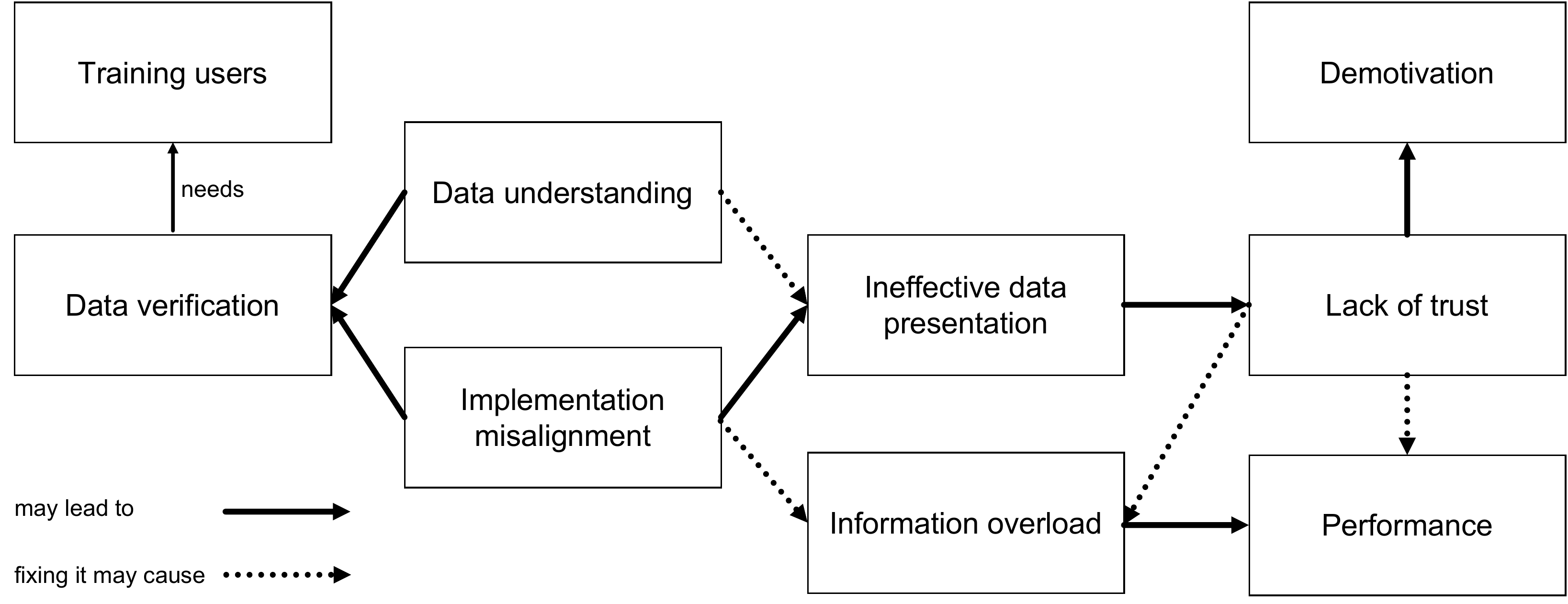}
\vspace{-0.5em}
\Description{The framework shows how problems interact (e.g., data understanding issue may lead to data verification issue, which then needs user training; also, fixing data understanding problems may cause ineffective data presentation.)}
\caption{A framework on the dependencies between challenges and possible side-effects of fixing problems.}
\vspace{-1.8em}
\end{figure*}


\looseness=-1
We have identified several development practices that can contribute to the aforementioned problems, such as the creativity in creating visualisations. Though, developers indicated that data visualisation innovations, although prevalent, are inadvisable as explained in Section~\ref{4.4}. The introduction of new visualisation artefacts may require the use of brief tutorials or even training. The major challenges reported by developers included data verification and data quality issues. A recurring source of conflict was shifting users' interest from legacy reporting applications such as spreadsheets to current dashboards~\citep{tory2021finding}. The above problems are just a sample of the most serious ones. Section~\ref{4} encompasses a plethora of diverse problems, some of which are already known in the information visualisation literature.

While most developers follow data visualisation guidelines when creating dashboards, the guidelines they use vary considerably. Many guideline sets are disjoint while some vendor-specific guidelines (e.g., Tableau \citep{Tableau2022}) take into account their own performance aspects and not only visual features. Results indicate a tension between generalistic user-interface guidelines and vendor-specific ones in that developers think that by using one guideline set, they address all the problems. Moreover, even though some developers use exclusively a single dashboarding tool, they do not stick to the guidelines published by the tool vendor as mentioned in Section~\ref{4.3}. Some developers do not follow guidelines because they need to cater for users' preferences that often go against these guidelines (e.g., choosing inappropriate visualisations). Also, addressing specific user needs in a particular dashboard may not be appropriate for another user, which suggests that tailoring dashboards is beneficial. 

The heavy reliance of developers on dashboarding tools determines the design decisions in that only those functionalities within the tools' repertoire can be included. When they can be included, there is little or even no room for extending them. This was clear on tailoring capabilities and user interaction data collection. While these tools enable developers to build dashboards responsively, they sacrifice the inclusion of features that may address their users' needs.

Dashboards are data-driven decision making tools, but they can also be used in an analytical, operational, or strategic manner, as opposed to other similar tools (e.g., self-reflection tools are more analytical \citep{choe2017understanding}). Though, developers did not focus on analytical features such as advanced data blending capabilities or sharing and exporting. In fact, developers were generally against exporting dashboard views as they believe dashboards are reporting tools. As prior research points out \citep{dimicco2016user}, following a user-centred approach to dashboard development can be challenging especially with opposing views on usability between users and developers \citep{hertzum2011personal}. The diversity of dashboard experts' backgrounds, development methodologies, subscribed visualisation guidelines, their data familiarisation habits, their dependence on proprietary dashboarding tools, and changing users requirements are all factors that lead to problematic dashboards.\\

\noindent\textbf{RQ2. What are the possible \emph{adaptation techniques} to the challenges? Can tailoring, especially \emph{automatic adaptations}, solve these challenges?}\\
Developers reported many adaptations to user challenges, which have been described in Section \ref{4} and summarised in Tables \ref{table:theme1table} to \ref{table:theme4table} in the Appendix. Many of the reported solutions to accommodate user needs and address barriers had been already employed in the participants' dashboards. However, some others were broad suggestions or currently non-actionable desires such as ``simplifying the views as much as possible'' for ineffective data presentation. Furthermore, some participants could not give any suggestions to raised challenges such as data verification and interoperability issues (as we asked for solutions to problems they had not reported). This highlights the need for further research to explore the feasibility of the adaptations. The developers' domains of expertise, although diverse, did not have a noticeable effect on their development practices or suggested adaptations. Though, developers emphasised the unique characteristics of their users and advised against generalising their views to other domains. \citet{sarikaya2018we} found that a dashboard's visual and functional aspects can reflect their application domain. However, our results suggest that relying on domain-specific features solely (e.g., common visualisation methods) to build dashboards may not be effective. This challenges previous research that places heavy focus on domain features to design dashboards~\citep{vazquez2019connecting, rojas2020cities}.

    
\paragraph{Can tailoring, especially \emph{automatic adaptations}, solve these challenges?}
Our findings confirm previous research that adaptive dashboards are the least used kind of tailored dashboards: eleven instances of dashboards incorporate customisation and/or personalisation functionalities, while four instances only use automatic adaptation~\cite{vazquez2019tailored}. It also confirms that tailoring dashboards to users' needs is advantageous and can increase their adoption and use~\citep{peischl2013can, sarikaya2018we, weggelaar2018developing, dabbebi2017towards, vazquez2020representing, eckerson2010performance}. However, our findings also indicate that personalisation is the most used type of tailoring on dashboards. This contradicts the work of \citet{vazquez2019tailored}, which found customisable dashboards to be the most common. Although developers are wary about including customisation by default and prefer to limit it to advanced users, we found customised dashboards to be still frequently used. Whereas customisation increases functional flexibility, it requires users to be aware of their needs. As a result, adaptive solutions can be more effective when users are not certain about their needs~\citep{vazquez2019tailored}. This is supported by the agreement on the applicability of the automatic adaptations presented to developers as described in Section~\ref{4.5}. Although developers were overwhelmingly in favour of automatic adaptations, most of them do not use it on their dashboards. This might be because they perceive obstacles in their technical implementation or were cautious on the effects of adaptive views on the users. The interviewees all agreed that the web-based adaptations shared with them were feasible, but these adaptive content presentations are only a subset of the adaptations for the Web available~\citep{brusilovski2007adaptive}. Other adaptations such as those based on social mechanisms (perceived community interests) or textual mechanisms (what to say and how to say it) are more applicable to other systems such as learning management or healthcare systems.

Tables \ref{table:theme1table} to \ref{table:theme4table} in the Appendix enumerate user challenges, corresponding adaptations (Section \ref{4}) and the type of tailoring that can be used to implement them. The distinction between the categories of tailoring highlights who the stakeholders are, and how, with each of these tailoring mechanisms, the degree of control of the system shifts between the end-user (customisation), the developer (personalisation) and the algorithm that models users (automatic adaptations). Customisable adaptations enable the user to change a dashboard's aspect: for example, allowing the user to drill down into raw data sets to mitigate users' lack of trust in the visualisation. Personalisable adaptations are any solution or features that can be given to a user \emph{at dashboard loading time} based on their role, goal, preferences or abilities to alleviate a challenge. For example, combining data into smaller subsets for a CEO to give a general overview and enabling drilling on demand. Finally, automatic adaptations are any adaptation that can be done \emph{at real-time} to change a dashboard aspect based on the same factors of personalisable adaptations in addition to data structure or sources~\citep{vazquez2019tailored}. Some adaptations can be done in any of the three forms such as enabling swapping between several charts. Other adaptations fit a specific type more such as the automatic adaptation of caching data seen by another similar user. Generally, adaptations that focus on the user are customisations whereas dashboard interface alterations performed by the system are either a personalisation or an automatic adaptation or both.

\subsection{Implications for Design}

\paragraph{Experiment with multiple forms of tailoring}
Tables \ref{table:theme1table} to \ref{table:theme4table} in the Appendix summarise actionable adaptations by developers. Yet, we propose the type of tailoring that can be used to implement these adaptations (e.g., allowing the user to zoom into the data set can be done automatically by the system or left to the user). Many of these adaptations can be performed automatically, while others such as `creating visualisations through an interactive dialogue' require user involvement. Tailoring via automatic adaptations could be convenient to users but it entails building user models that contain features of the users that are relevant for these purposes (e.g., user goal). For instance, we learned that users employ workarounds when encountering problems in dashboards (e.g., looking at raw data when the functionalities do not allow to narrow down the data set). These strategies are indicators of problems so automated adaptations could be implemented when these workarounds are detected. Since developers are not very keen on customisation, starting with system-based tailoring (personalisation and automatic adaptation) may be a better approach. Customisation functionalities can be introduced gradually when the user has acquired the needed competency. It is also important to investigate how users respond to each kind of tailoring, ideally through unobtrusive usage data collection. 

\paragraph{Tailor dashboards to help developers, too}
There were several problems of major concern to developers. For example, training dashboard users, a problem described earlier as \emph{huge} \citep{tendedez2018scoping}, was reported often by the participants. Training can be built within the dashboard as a course tailored to the users' skills, or expandable tips and tutorials that pop up when users have a problem. Also, developers' efforts to make dashboards that answer all users' questions fall short. These overheads could be saved if one dashboard was made and then tailored to each user needs. Tailoring can target dashboard information, visualisation types, layout, functionalities and interaction. It can be done based on users' roles, goals, preferences and skills~\citep{vazquez2019tailored}. Tailoring can also ease developers' efforts to move users from traditional reporting systems to dashboards. A dashboard's complexity can evolve along with the user's visual literacy starting with simple charts to more complex ones~\citep{froese2016lessons}. During this process, and to increase users' motivation, the dashboard could highlight its strengths and advantages compared to spreadsheet software (e.g.,\ by showing analytics like time saved).

\paragraph{Put users in control of their own dashboards}
\looseness=-1
Although developers were generally wary of giving users full customisation control over their dashboards, enabling customisation has potential to address several issues. When users are unable to describe what they need, they could still be able to envisage it. Customisation will give users the chance to create their own charts on demand. Feedback on the created charts can be supplied to users within the dashboard if mistakes were made or if charts, or the dashboard as a whole, could be improved. This also has the potential to increase users' visual literacy by learning from the feedback on their own mistakes. Moreover, the dashboard system can learn from these customisations by recommending visualisations when similar users demand similar data. Another area of improvement that can be addressed with customisation is visual design. For example, developers suggested  that some users struggled with something as widely agreed-upon such as the traffic-light colour coding. Customisation could easily solve this problem for users, and the dashboard system could create automatic adaptations informed by user choices.

\paragraph{Leverage asynchronous communication to fasten development}
\looseness=-1
Agile software development methodologies are typically used to address evolving user needs and constant changes~\citep{alvertis2016user}. However, as made clear by the participants, agile sprints are not fast enough to keep up with the demands of dashboard development. This is because data nowadays is released at an unprecedented rate, which can quickly change requirements. Also, dashboards are relatively new as an information presentation tool, so users are learning along the way and their realisation of their needs evolves accordingly. By the time for the next agile sprint, users acquire new knowledge rendering their previous requirements obsolete. One solution may be a shared asynchronous communication method in a participatory design approach \citep{gilliot2018participatory, martins2017development}: developers could share instantaneous updates of the dashboards, and users could see the product before it is completely implemented. At their preferred time, users could provide early feedback to change a required visualisation if they think it no longer satisfies their needs. Developers could then attend to users' feedback earlier than they would with periodic agile meetings. Users could also share rough sketches of their needs or screenshots of similar visualisations or dashboards they would like to have. It could also be a better way for eliciting all users' needs when gathering users and agreeing on universal requirements proves difficult. Such a solution can enable shifting the course of development rapidly with less cost on the developer's side.

\paragraph{Create a channel for remote support}
The COVID-19 pandemic has driven rapid adoption of online environments and dashboard development can make use of this change. As dashboard users were commonly found to request service (help completing a task) and assistance  \citep{tory2021finding}, developers could provide technical support within the dashboard through screen sharing (access to users' screen) or through configuration sharing. An example of this technique was given by a participant where data filters are applied on the developer side then the customised view is shared with the user. Configuration can also be shared between users by extending the customisation feature so that they can learn from each other. This way, developers will have a better idea about the specific problems users encounter instead of re-actively responding to them \emph{if} they are reported, or if users are visited later.
\section{Conclusion}
We explore dashboard developers' perspectives on the users and their interactions with dashboards. Our findings show that developers are aware of many problems users encounter such as ineffective data presentations and those problems caused by gaps in visual literacy. We also learned about some developers practices that contribute to existing problems. Encouragingly, our findings also indicate that developers implement mechanisms to overcome these challenges with user-specific adaptations. Such adaptations constitute an avenue for future research and practice.

\bibliographystyle{ACM-Reference-Format}
\bibliography{references}

\appendix
\section{Challenges, Adaptations, and Tailoring Forms}
\begin{table*}[b]
    \Description{Implementation misalignment problem can be adapted by giving users a starting point or conducting a workshop to understand their needs. These adaptations are implemented and can be done using personalisation or auto adaptation.}
  \caption{User challenges from `Involving Users in the Development' theme (first column), adaptations to each challenge \emph{given by participants} (second column), whether it is implemented (I) or suggested (S) (third column) and the kind of tailoring \emph{suggested by the authors} to implement the adaptation (C=Customisation, P=Personalisation and AA=Automatic Adaptation).
  \label{table:theme1table}}
  \centering
  \begin{tabular}{p{0.18\textwidth}p{0.57\textwidth}p{0.05\textwidth}p{0.05\textwidth}p{0.05\textwidth}p{0.05\textwidth}}
  \toprule
    \multirow{2}{*}{\textbf{Challenge}} & \multirow{2}{*}{\textbf{Adaptation}} & \multirow{2}{*}{\textbf{I/S}} & \multicolumn{3}{c}{\textbf{Type of Tailoring}}\\
    & & & \textbf{C} & \textbf{P} & \textbf{AA}\\
    \midrule
    \multirow{2}{*}{\begin{tabular}[t]{@{}c@{}}Implementation\\Misalignment\end{tabular}}
    & - Giving users a starting point to facilitate requirement elicitation. & I & & x & x \\
    & - Conducting a workshop to understand users' available tools. & I & & x & x \\
    \midrule
    \multicolumn{3}{c}{\textbf{Sum of adaptations}} & \textbf{0} & \textbf{2} & \textbf{2}\\
    \bottomrule
  \end{tabular}
  \label{table:adaptationSummary2}
\end{table*}

\begin{table*}[b]
    \Description{Example of issues & adaptations in the second theme: data quality issue is addressed by removing erroneous or incomplete data (implemented) or by using dedicated tools to handle heterogeneous data (suggested). Both can be implemented using auto-adaptation.}
  \caption{User challenges from `All about Data: Performance, Access, Provenance, Currency and Metadata' theme (first column), adaptations to each challenge \emph{given by participants} (second column), whether it is implemented (I) or suggested (S) (third column) and the kind of tailoring \emph{suggested by the authors} to implement the adaptation (C=Customisation, P=Personalisation and AA=Automatic Adaptation).
  \label{table:theme2table}}
  \centering
  \begin{tabular}{p{0.18\textwidth}p{0.57\textwidth}p{0.05\textwidth}p{0.05\textwidth}p{0.05\textwidth}p{0.05\textwidth}}
  \toprule
    \multirow{2}{*}{\textbf{Challenge}} & \multirow{2}{*}{\textbf{Adaptation}} & \multirow{2}{*}{\textbf{I/S}} & \multicolumn{3}{c}{\textbf{Type of Tailoring}}\\
    & & & \textbf{C} & \textbf{P} & \textbf{AA}\\
    \midrule
    \multirow{2}{*}{Data Quality Issues}
    & - Removing erroneous or incomplete data elements before showing data. & I & & & x \\
    & - Using dedicated tools to handle heterogeneous data. & S & & & x \\
    \midrule
    \multirow{4}{*}{\begin{tabular}[t]{@{}c@{}}Dashboard\\Performance\end{tabular}} 
    & - Employing data caching techniques. & I & & & x \\
    & - Loading in background closely related data and display when needed. & S & & & x \\
    & - Showing data extract (a compressed version). & I & & & x \\
    & - Refreshing data displayed periodically instead of live data. & I & & & x \\
    \midrule
    \multirow{2}{*}{Lack of Trust} 
    & - Showing data source and metadata. & I & x & x & x \\
    & - Allowing the user to zoom into the data set. & S & x & x & x \\
    \midrule
    Protecting\\Data Access
    & - Using workflows connecting data sources to user roles. & I & & x & x \\
    \midrule
    \multicolumn{3}{c}{\textbf{Sum of adaptations}} & \textbf{2} & \textbf{3} & \textbf{9}\\
    \bottomrule
  \end{tabular}
  \label{table:adaptationSummary2}
\end{table*}

\begin{table*}[b]
    \Description{Example of issues & adaptations in the third theme: information overload can be adapted by combining data into smaller subsets & enabling drill-down on demand. This is a suggested adaptation that can be done in one of the three tailoring techniques.}
  \caption{User challenges from `The Tensions of Addressing User Needs' theme (first column), adaptations to each challenge \emph{given by participants} (second column), whether it is implemented (I) or suggested (S) (third column) and the kind of tailoring \emph{suggested by the authors} to implement the adaptation (C=Customisation, P=Personalisation and AA=Automatic Adaptation).
  \label{table:theme3table}}
  \centering
  \begin{tabular}{p{0.18\textwidth}p{0.57\textwidth}p{0.05\textwidth}p{0.05\textwidth}p{0.05\textwidth}p{0.05\textwidth}}
  \toprule
    \multirow{2}{*}{\textbf{Challenge}} & \multirow{2}{*}{\textbf{Adaptation}} & \multirow{2}{*}{\textbf{I/S}} & \multicolumn{3}{c}{\textbf{Type of Tailoring}}\\
    & & & \textbf{C} & \textbf{P} & \textbf{AA}\\
    \midrule
    \multirow{3}{*}{Information Overload} 
    & - Combining data into smaller subsets \& enabling drill down on demand. & S & x & x & x \\
    & - When several users use the same dashboard, involving users to keep relevant charts and remove the rest. & I & x & x & x \\
    & - Educating users on the risks of displaying too much information. & I & & x & x \\
    \midrule
    \multirow{6}{*}{\begin{tabular}[t]{@{}c@{}}Ineffective Data\\ Presentation\end{tabular}}
    & - Enabling users to change visualisation type on demand. & I & x & &
    \\
    & - Using data visualisation guidelines. & I & & x & x \\
    & - Creating visualisations informed by the questions that the dashboard is trying to answer. & I & & x & x \\
    & - Creating visualisations through an interactive dialogue. & I & x & & \\
    & - Creating visualisations based on KPIs to be represented. & I & x & & \\
    & - Using the same colour for the same kind of data across dashboard(s). & I & x & x & x \\
    \midrule
    \multicolumn{3}{c}{\textbf{Sum of adaptations}} & \textbf{6} & \textbf{6} & \textbf{6}\\
    \bottomrule
  \end{tabular}
  \label{table:adaptationSummary2}
\end{table*}

\begin{table*}[b]
    \Description{Example of issues & adaptations in the fourth theme: Demotivation can be addressed by showing users suggestions to improve their performance or presenting information based on what motivates them. The former adaptation is implemented while the latter is suggested; both can be done with personalisation or auto-adaptation.}
  \caption{User challenges from `Adoption, Onboarding and Facilitating the Learning Journey' theme (first column), adaptations to each challenge \emph{given by participants} (second column), whether it is implemented (I) or suggested (S) (third column) and the kind of tailoring \emph{suggested by the authors} to implement the adaptation (C=Customisation, P=Personalisation and AA=Automatic Adaptation).
  \label{table:theme4table}}
  \centering
  \begin{tabular}{p{0.18\textwidth}p{0.57\textwidth}p{0.05\textwidth}p{0.05\textwidth}p{0.05\textwidth}p{0.05\textwidth}}
  \toprule
    \multirow{2}{*}{\textbf{Challenge}} & \multirow{2}{*}{\textbf{Adaptation}} & \multirow{2}{*}{\textbf{I/S}} & \multicolumn{3}{c}{\textbf{Type of Tailoring}}\\
    & & & \textbf{C} & \textbf{P} & \textbf{AA}\\
    \midrule
    \multirow{2}{*}{Demotivation} 
    & - Showing users suggestions to improve their performance. & I & & x & x \\
    & - Presenting information based on what motivates the user. & S & & x & x \\
    \midrule
    \multirow{7}{*}{User Training Issues}
    & - Training users in hand holding period or in formal training sessions. & I & & x & x \\
    & - Putting users into groups to introduce them to their dashboards separately. & I & & x & \\
    & - Showing instructions on using the dashboard via pop-up messages, annotations, videos and tutorials. & I & & x & x \\
    & - Emphasising capabilities and limitations of dashboard to eliminate confusion. & I & & x & x \\
    & - Showing suggestions and asking for feedback to enhance training. & I & & x & x \\
    & - Showing tutorials to first-time users or who have not logged in for a while. & I & & x & x \\
    & - Educating users when asking for inappropriate visualisations. & I & & x & x \\
    \midrule
    \multirow{4}{*}{\begin{tabular}[t]{@{}c@{}}Functionality Use\\Issues\end{tabular}} 
    & - Having a mechanism to apply filters on dashboards then sharing them with users. & I & & x & x \\
    & - Changing filters to fit user's role needs. & S & & x & \\
    & - Displaying a video of how to use functionalities and navigate the dashboard. & I & & x & x \\
    & - Highlighting interesting areas or where a user should click. & I & & x & x \\
    \midrule
    \multirow{7}{*}{\begin{tabular}[t]{@{}c@{}}Data Understanding\\ Issues\end{tabular}}
    & - Going through the interpretation of the visualisations explicitly. & I & & x & x \\
    & - Adding extra helping information (e.g. axes labels). & I & & x & x \\
    & - Adding data points to give context to maps and terms. & I & & x & x \\
    & - Adding legends even for familiar colour palettes. & I & & x & x \\
    & - Using the least number of charts (maximum two by default). & S & x & x & x \\
    & - Utilising rules associated with charts to highlight changes automatically. & I & & x & x \\
    & - Including definitions of KPIs of a variety of industries/domains. & I & x & x & x \\
    \midrule
    \multirow{6}{*}{\begin{tabular}[t]{@{}c@{}}Inappropriate\\ Information Placement\end{tabular}}
    & - Using white space to emphasise groups of charts. & I & & & x \\
    & - Keeping common functionalities in a particular area. & I & & x & x \\
    & - Enabling sorting dashboard content to explore different outcomes. & S & x & & \\
    & - Creating responsive layouts to be used on multiple devices. & I & & & x \\
    & - Organising charts in the order they should be read. & I & x & x & x \\
    & - Grouping dashboards by type to ease navigation. & I & x & x & x \\
    \midrule
    
    \multicolumn{3}{c}{\textbf{Sum of adaptations}} & \textbf{5} & \textbf{23} & \textbf{23}\\
    \bottomrule
  \end{tabular}
  \label{table:adaptationSummary2}
\end{table*}

\end{document}